\documentclass[10pt, aps,prc,twocolumn,superscriptaddress,preprintnumbers,
amsmath, 
floatfix,
longbibliography,
nofootinbib
]{revtex4-1}
\usepackage[T1]{fontenc}
\usepackage[utf8]{inputenc}
\usepackage{adjustbox}          
\usepackage[caption=false]{subfig}
\usepackage{url}
\usepackage{color}
\usepackage{float}
\usepackage[pdftex,colorlinks=true, linkcolor = blue, citecolor=blue,urlcolor=blue, bookmarksnumbered=true, bookmarksopen=true]{hyperref}
\usepackage{longtable}
\usepackage{amsfonts}
\usepackage{dsfont}
\usepackage{wrapfig,bm} 
\usepackage[normalem]{ulem}
\usepackage{MnSymbol}
\usepackage{float}
\usepackage{color,soul}
\newcommand{\beq}{\begin{equation}}
\newcommand{\eeq}{\end{equation}}
\newcommand{\bea}{\begin{eqnarray}}
\newcommand{\eea}{\end{eqnarray}}

\begin{document}
\title{Neck Rupture and Scission Neutrons in Nuclear Fission}


\author{{Ibrahim Abdurrahman}} 
\affiliation{Theoretical Division, Los Alamos National Laboratory, Los Alamos, NM 87545, USA}   
\author{{Matthew Kafker}}
\affiliation{Department of Physics, University of Washington, Seattle, WA 98195--1560, USA}
\author{{Aurel Bulgac}}
\affiliation{Department of Physics, University of Washington, Seattle, WA 98195--1560, USA}
\author{{Ionel Stetcu}}
\affiliation{Theoretical Division, Los Alamos National Laboratory, Los Alamos, NM 87545, USA}   


   
\date{\today}

\begin{abstract}

Just before a nucleus undergoes fission, a neck is formed between the emerging fission fragments. It is widely accepted that this neck undergoes a rather violent rupture, despite the absence of unambiguous experimental evidence. 
The main difficulty in addressing the neck rupture and saddle-to-scission stages of fission is that both are highly non-equilibrium processes. Here, we present the first fully microscopic characterization of the scission mechanism, along with the spectrum and the spatial distribution of scission neutrons (SNs), and some upper limit estimates for the emission of charged particles. The spectrum of SNs has a distinct angular distribution, with neutrons emitted in roughly equal numbers in the equatorial plane and along the fission axis.  They carry an average energy around $ 3 \pm 0.5 $ MeV for the fission of $^{236}$U, $^{240}$Pu and $^{252}$Cf, and a maximum of 16 - 18 MeV. We estimate a conservative lower bound of $9-14$ \% of the total emitted neutrons are produced at scission.

\end{abstract}  

\preprint{NT@UW-23-08,LA-UR-23-25884}

\maketitle


Nuclear fission was experimentally discovered by \textcite{Hahn:1939} in 1939.  Later in 1939, it was named and its main mechanism was explained by \textcite{Meitner:1939}.  It is a quantum many-body process of extreme complexity, with various parts of the process occurring at vastly different timescales.
The total time it takes, from the moment a neutron initiates the formation of a compound nucleus 
until all final fission products have attained their equilibrium state after $\beta$-decay, 
 can be on the order of billions of years \cite{Gonnenwein:2014}, and is greater by enormous orders of magnitude relative to the time it takes a nucleon to cross a nucleus, ${\cal O}(10^{-22})$ sec.

The compound system, formed by a
low-energy neutron~\cite{Hahn:1939} interacting with a target nucleus, evolves through many distinct stages. The first stage is a relatively slow quasi-equilibrium evolution, that lasts until the compound 
system~\cite{Bohr:1936} 
reaches the outer saddle-point at $\approx 10^{-14}$ sec.~\cite{Gonnenwein:2014}. During this stage, the nucleus, with an initial prolate intrinsic shape and axial symmetry, evolves into a nucleus with triaxial shape, and eventually into a reflection asymmetric and axially symmetric elongated shape near the outer fission barrier~\cite{Ryssens:2015}.
The second stage is a highly non-equilibrium evolution from 
saddle-to-scission~\cite{Bulgac:2016,Bulgac:2019c,Bulgac:2020}, when the primordial fission fragments (FFs) properties are defined within a duration of $\approx 5\times10^{-21}$ sec.~\cite{Gonnenwein:2014}. Even though this second stage is much faster than the first stage, it corresponds to rather slow dynamics, relative to the third stage (scission).
In this stage, the compound nucleus undergoes a relatively rapid separation into two FFs, lasting $\approx 10^{-22}$ sec. This stage is also known in the literature as the  neck rupture.  
This is followed by a fourth stage, the  FFs Coulomb acceleration during an interval of time of ${\cal O}(10^{-18})$ sec., when at the end the FFs achieve 
a shape equilibration.
While the initial compound nucleus is a relatively cold system with a very small spin, the primordial FFs are very hot and have relatively large spins as well \cite{Bulgac:2019c,Bulgac:2021}. 
These highly excited FFs emit prompt neutrons for an interval of time up to about ${\cal O}(10^{-14})$ sec., followed by the emission of the majority of
prompt $\gamma$-rays  for an interval of time until about ${\cal O}(10^{-3})$ sec., which is further followed by much slower $\beta$-decays. 
Other processes are also possible, such as delayed neutron or $\gamma$ emission after $\beta$-decay. 

With the exception of the saddle-to-scission configuration and the neck rupture, all the other stages of fission are relatively slow quasi-equilibrium processes. 
The fission dynamics after the compound nucleus reaches the outer saddle point has been typically
described in terms of the potential energy surface of the nucleus, determined by its shape~\cite{Gonnenwein:2014,Ring:2004,Pomorski:2012,Schunck:2016},
not compatible with recent microscopic studies~\cite{Bulgac:2016,Bulgac:2019c,Bulgac:2020, Bender:2020}, and agreement with experimental data typically requires adjustment of  many parameters \cite{Sierk:2017}. On the other hand, many different approaches to FF mass and charge distributions lead to agreement with experiment~\cite{Verriere:2021,Mumpower:2020,Sierk:2017,Ivanyuk:2024,Sadhukhan:2016,Sadhukhan:2017}, even though they rely on clearly contradicting physics assumptions, which simply demonstrates that these distributions are not very sensitive measures of the fission dynamics.  
Approximately at the top of the outer saddle, the nucleus starts forming a barely seen ``wrinkle'', where eventually the neck between the two FFs is formed. 
This ``wrinkle'' tends to appear when the fissioning nucleus is at a very early stage during the descent to scission, and its position hardly changes in time. A significant change in this position would require a large amount of energy for displacement that would not be available from fluctuations \cite{Hill:1953,Griffin:1957,Reinhard:1983,Bulgac:2019d,Bulgac:2022}. At the top of the outer saddle the nucleus starts a relatively slow dissipative evolution 
towards scission~\cite{Bulgac:2016,Bulgac:2019c,Bulgac:2020}.  During this period, the fissioning nucleus 
gets more elongated and the neck becomes more and more pronounced. 
The nuclear fluid behaves as nuclear molasses, with a very small collective velocity~\cite{Bulgac:2016,Bulgac:2019c,Bulgac:2020},  while at the same time the
intrinsic temperature of the system gradually increases. The bond between the two fission partners slowly weakens until the neck, which was still keeping them together, reaches 
a critical small diameter of approximately 3 fm and ruptures, exactly where the initial ``wrinkle'' formed much earlier at the top of the outer saddle. 
This dramatic separation of the two emerging FFs is  a rather short-time event. 
For \textcite{Brosa:1990} scission was the defining stage of fission, where the total kinetic energy (TKE) of the FFs is defined along with the average FF properties. 
The Brosa model assumes that the nucleus is a very viscous fluid, with a long neck that ruptures 
at a random position, and is widely invoked today in many phenomenological models \cite{Oberstedt:1999,Vogt:2009,Becker:2013,Talou:2021,Litaize:2012,Lovell2021,Fujio:2023,Najumunnisa:2023}, even though it has no microscopic justification and its claimed grounding in experimental data does not necessarily support a unique interpretation. Additionally, the Brosa random neck rupture model contradicts the theoretical assumptions of 
other popular approaches, such as the scission-point model of \textcite{Wilkins:1976}, 
where the FF formation is based on statistical equilibrium~\cite{Lemaitre:2015,Lemaitre:2021}, and Brownian motion or Langevin models~\cite{Sierk:2017,Randrup:2011,Albertsson:2020,Verriere:2021a,Ivanyuk:2024}. 
The drama of scission is followed by unavoidable debris characteristic of such dramatic separations, 
the scission neutrons (SNs), envisioned as early as 1939 by \textcite{Bohr:1939}.  Potentially other heavier fragments,
usually termed as ternary fission products~\cite{Vandenbosch:1973, Rose:1984, Wagemans:1991}, are created as well.  We relegate a  brief review of the history of SNs as an online supplementary material \cite{supplement}, with additional references~\cite{Debenedetti:1948,Fraser:1952,Fraser:1954,Stavinsky:1959,Halpern:1959,Vorobyev:2010,Brosa:1992,Rizea:2008,Stetcu:2011,Tanimura:2017,Ren:2022,Scamps:2012,Bulgac:2023,Demers:1946,Farwell:1947,Wollan:1947,Marshall:1949,Titterton:1951,Allen:1950,Feather:1947,Nobles:1962,Fraenkel:1967,Albenesius:1959,Watson:1961,Marshall:1966,Goward:1949,Whetstone:1965,Dakowski:1967,Cosper:1967,Krogulski:1969,Chwaszczewska:1967,Vorobiev:1969,Vandenbosch:1973,Rose:1984,Poenaru:1986,Dalfovo:1999,Pethick:2008,Giorgini:2008,Fowler:1928,Uehling:1933,Bertsch:1988,Bulgac:2023b,Tully:1971,Tully:1990,Hammes:1994,Frobrich:1998}, where we also present many more details of our study. 

\begin{figure}
    \centering
    \includegraphics[width=0.4 \textwidth, clip]{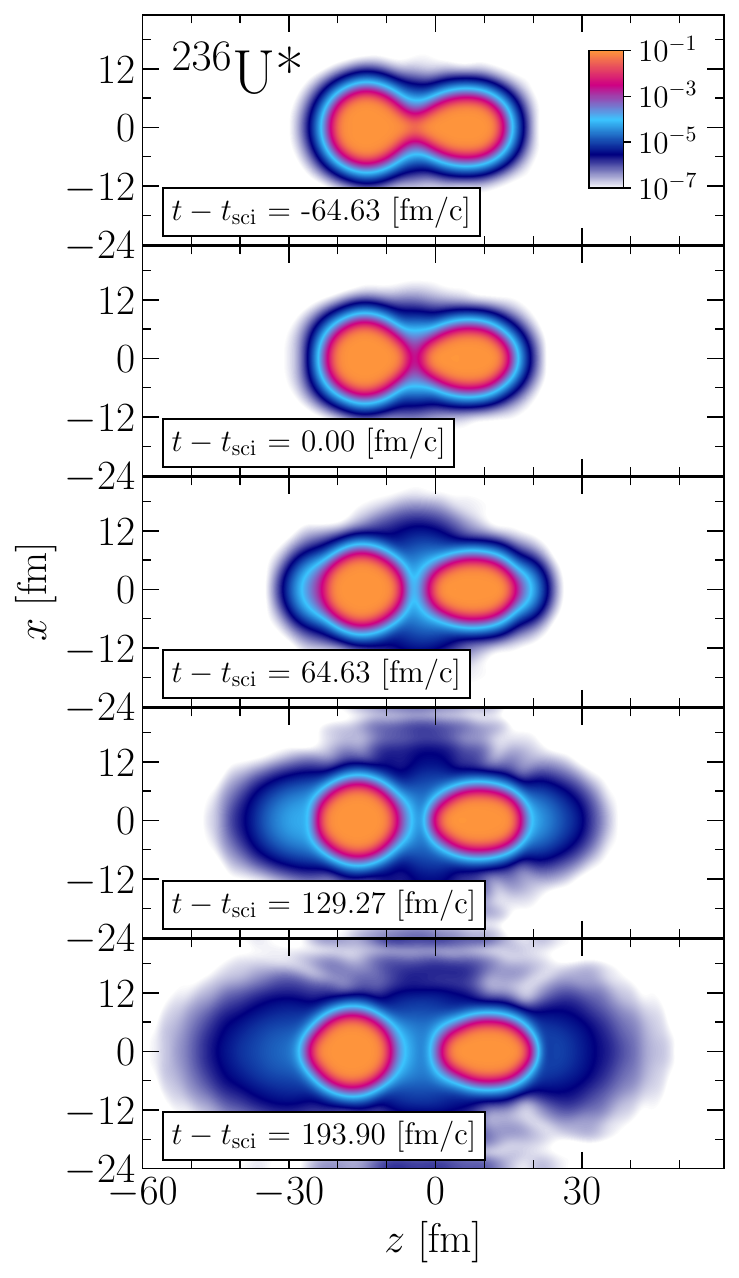}
    \caption{Time series of the neutron number density in fm$^{-3}$ for a typical fission trajectory.  Similar results for protons are contained in the supplementary material \cite{supplement}.}
    \label{fig:tseries}
\end{figure}

\begin{figure}
    \centering
   \includegraphics[width=1 \columnwidth]{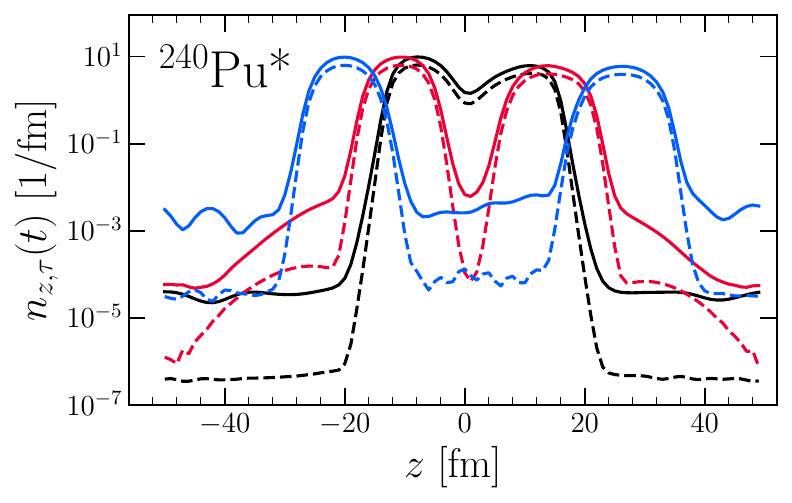}
    \includegraphics[width=1 \columnwidth]{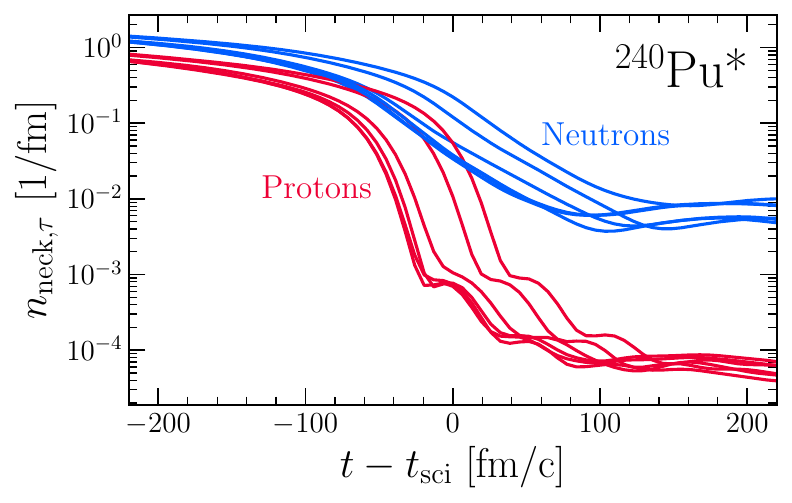}
    \caption{
   In the upper panel we display the integrated nucleon density along the fission axis $ n_{z,\tau}(t) = \int dxdy \, n_\tau(x,y,z,t)$ at several times; before scission at -258.53 fm/c (black lines), when the neck is barely formed; after scission at 129.27 fm/c (red lines); and after the FFs separated respectively at 517.06 fm/c (blue lines). Neutrons/protons are represented via solid/dashed lines respectively. 
    In the bottom panel, the nucleon number density integrated over the cross section of the neck as a function of time. A fit around the scission time,  shows that integrated over the cross section of the neck neutron number density decay exponentially, 
   $ n_{\mathrm{neck},\tau}(t) \sim \exp(-t/\tau)$, with $\tau \approx 35.0 \pm 2.2$ fm/c for neutrons and $15.3\pm 0.3$ fm/c for protons. }
    \label{fig:neck} 
    \end{figure}

\begin{figure}
    \centering
     \includegraphics[width=1.0 \columnwidth]{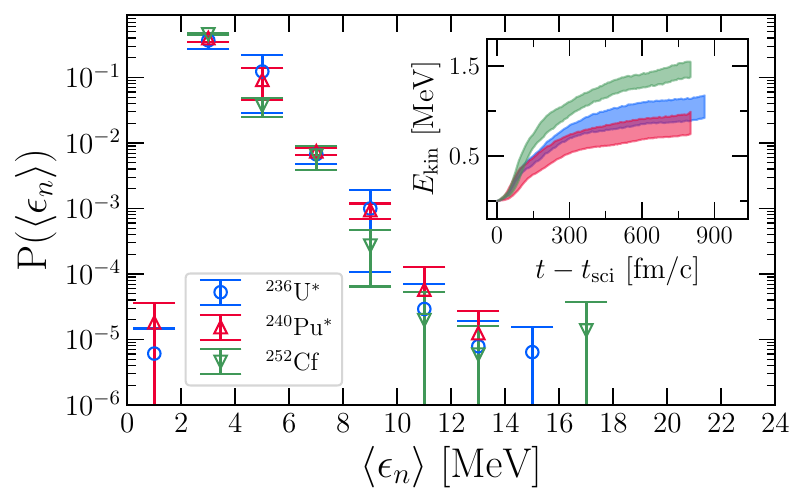}

   \caption{The scission neutron kinetic energy distribution with uncertainties corresponding to different trajectories.  The distributions are normalized to the total number of SNs i.e. $\sum \mathrm{P} ( \langle \epsilon_n \rangle ) \times \frac{\Delta E} {\mathrm{MeV}} = N_{\mathrm{sci}} $, with $\Delta E = 2$ MeV. The inset shows the total kinetic energy of the SNs vs time. The shaded regions represent the standard deviation from considering various trajectories.
   }
    \label{fig:ke}
\end{figure}
%

\begin{figure}
    \centering
    \includegraphics[width=1.0 \columnwidth]{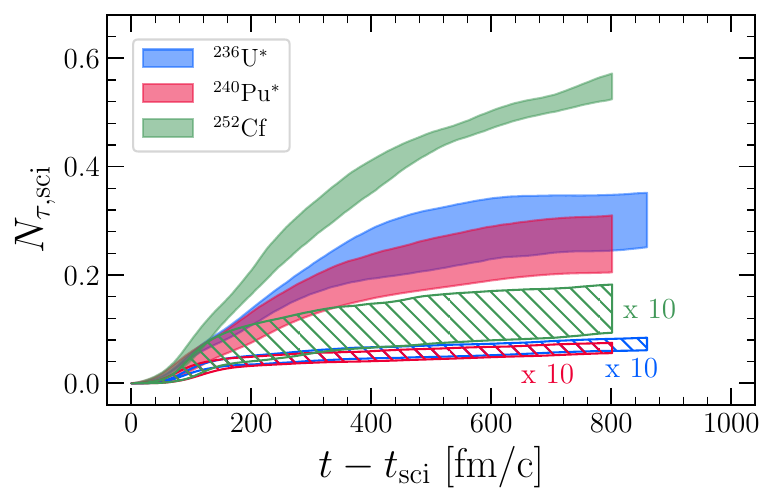}    
    \caption{The solid/dashed regions show the number of SNs/protons respectively; $\tau = n,p$.
    The shaded regions represent the standard deviation from considering various trajectories.  $N_{p,sci}$ was enhanced by a factor of 10. }
    \label{fig:nsci}
\end{figure}
%


In these simulations, we started by placing the initial compound nucleus 
near the top of the outer barrier in a very large simulation volume, in order to allow to the emitted nucleons enough time to decouple from the FFs 
after the neck rupture. 
We have performed a range of simulations for $^{235}$U(n$_{th}$,f), $^{239}$Pu(n$_{th}$,f), and $^{252}$Cf(sf),
using the nuclear energy density functional (NEDF)  SeaLL1~\cite{Bulgac:2018} in simulation volumes $48^2\times120$  and $48^2\times100$ fm$^3$, 
with a lattice constant of 1 fm, for further technical details see Ref.~\cite{Shi:2020}. 
The SeaLL1 NEDF is defined by only 8 basic nuclear parameters, each related to specific nuclear properties known for decades, and contains the smallest number of phenomenological parameters of any NEDF to date ~\cite{Bulgac:2018,Bulgac:2022c}.
We started the simulations at various deformations $Q_{20}$ and $Q_{30}$, as listed in Ref.~\cite{supplement}, near 
the outer fission barrier rim and see Refs.~\cite{Bulgac:2016,Bulgac:2019c,Bulgac:2020}, where 
one can find more details about how the FF properties vary with the choice of initial conditions. 
Our simulation volume of $48^2\times120$ fm$^3$ required the use of the entire supercomputer Summit (27,648 GPUs), corresponding to 442 TBs of total GPU memory, with further details provided in Ref.~\cite{supplement}. Despite this, we still could not follow the emission of nucleons for a long time, 
since the emitted nucleons are reflected back at the boundary relatively rapidly, see the lowest two rows of Fig.~\ref{fig:tseries}, where interference patterns emerge. In the transversal direction the reflection from the boundaries occurs 
earlier than along the fission axis, and that has effected some of the properties of the nucleons emitted perpendicular to the fission axis. However the effect is minor, see Ref.~\cite{supplement}.

From here, we will concentrate on the dynamics of the neck formation and rupture, followed by the emission of nucleons,   all treated   
within the time-dependent density functional theory extended to superfluid fermionic systems~\cite{Bulgac:2019}. The integrated neck density, shown in Fig.~\ref{fig:neck}, is defined as
\begin{align}
 n_{\mathrm{neck},\tau}(t) = \int \!\!dxdy \, n_\tau(x,y,z_\mathrm{neck},t), \quad \tau = n, \, p,
\end{align}
separately for neutrons and protons, where the $z_{neck}$ is the position along the fission axis $Oz$ where the neck has the smallest radius.  The neck decays relatively slowly at scission, until its diameter reaches about 3 fm, after-which it undergoes a very rapid decay.  Different curves illustrated in the lower panel correspond to trajectories started at various initial conditions for the deformations $Q_{20}, Q_{30}$ close to the outer fission barrier~\cite{supplement}. The time to reach scission can vary significantly, depending on the initial values of the deformations $Q_{20}, Q_{30}$ and on the NEDF used, typically ranging from 1,000 to 3,000 fm/c.

These microscopic results illustrate several points, which were unknown until now, due to the absence 
of any detailed fully microscopic quantum many-body simulations of fission dynamics.  First, the ``wrinkle'' in the nuclear density, where the neck is eventually formed and where the nucleus eventually scissions, is determined a long time before the nucleus reaches scission. Within the TDDFT framework the position of the neck rupture is not random,
unlike in the Brosa model~\cite{Brosa:1990,supplement}. At the time when the neck reaches a critical diameter of $\approx 3 $ fm, the nuclear surface tension and the shape of the compound around the neck region, can no longer counteract the strong Coulomb repulsion between the preformed FFs, causing the system to violently ``snap''. 
One should keep in mind that as the intrinsic temperature of the compound nucleus increases, the surface tension also decreases.    
The geometry of the nuclear shape changes dramatically at this stage, from exhibiting a neck region where the Gaussian curvature is negative, to two separated FFs with 
surfaces characterized by predominantly positive Gaussian curvatures. 

Second, the proton neck completes its rupture earlier than the neutron neck does, see lower panel in
 Fig.~\ref{fig:neck}, resulting in the neck being mostly sustained by the neutrons just before the full rupture.  This is similar to the neutron density in the neutron skin of heavy nuclei. In this time interval, the number of neutrons per unit area at the neck varies by an order of magnitude.
The protons in the emerging FFs separate about 50-100 fm/c before the neutrons neck ruptures. 
Additionally, the integrated neutron and proton densities at $z_{\mathrm{neck}}$ asymptotically reach almost equilibrium values, after the neck ruptures.

 Third, the rupture is unarguably the fastest stage of the fission dynamics, starting from the capture of the incident neutron and formation of the compound nucleus, until all fission products have been emitted. The decay times are 15 fm/c and 35 fm/c for proton and neutron necks respectively, which are significantly faster processes than the time it takes the fastest nucleon to communicate any information or facilitate any kind of equilibrium between the two preformed FFs, which is at minimum $ \sim 160$ fm/c. 

 Fourth, the neck decay dynamics displays a clear universality (for asymmetric fission) irrespective of the initial conditions, as shown in the lower panel of Fig.~\ref{fig:neck}, with the proton neck rupturing well ahead of the neutron neck, due to the presence of a well-defined neutron skin.  The time to full mass, charge, excitation energies and TKE definition all vary~\cite{Bulgac:2019c,Bulgac:2020}, while the neck dynamics are essentially unchanged for both proton and neutron components.

Last, the scission mechanism emerging from a fully microscopic treatment of the fission dynamics is totally at odds with previous models, including the Brosa random rupture model and the scission-point models. TDDFT extended to superfluid systems is the only theoretical microscopic framework so far in the literature in which scission is treated without any unchecked assumptions or fitting parameters, which produces results that are in agreement with data~\cite{Bulgac:2016,Bulgac:2019c,Bulgac:2020,Bulgac:2019}.



The neck rupture is a very fast ``healing'' process of the nuclear surface in the neck region.  Unlike a gas in a punctured balloon, which would rapidly escape the enclosure, 
due to the presence of the nuclear ``skin'' and strong surface tension, the nucleus behaves as a fluid. 
The  surface tension quickly ``heals'' the ``wound,'' however a small fraction of matter manages to escape like a gas, with no droplet formation,
see Fig.~\ref{fig:tseries}. The potential condensation of this emitted gas into light charged particles cannot be described within the present framework, which includes at most two particle correlations. This is not to be confused with ternary fission of a preformed fragment, where further discussion is provided in Ref.~\cite{supplement}.  In Fig.~\ref{fig:tseries}, and more in Ref.~\cite{supplement}, we show several 
representative frames for neutrons and protons of the neck formation and emission of nucleons. 

The most remarkable features of this process are the following.
 As visible in Fig.~\ref{fig:neck} the proton neck completes its rupture before the neutron neck, however in two stages.  
 Immediately at scission, which in Fig.~\ref{fig:tseries} is identified with the rupture of the proton neck, 
 a number of nucleons are emitted in the plane perpendicular to the fission axis (see second time frame).  After a sufficient time for nucleons to propagate from the neck to the nose of each FF, scission nucleons also appear propagating in front of each FF.  In 1984, it was suggested by \textcite{Madler:1985}, that the reabsorption of the neck stumps by the FFs, being a relatively rapid process, could act as a ``catapult'', which is more appropriately described as a slingshot, and ``push'' nucleons out of the front of the FFs.  It is also important to note, the formation of three neutron clouds, two in front of the FFs and one transverse ring perpendicular to the fission axis, appears across all considered trajectories and nuclei thus far.  

 Remembering that the TKE is roughly 171-186 MeV, at the end of the full Coulomb acceleration the light and heavy FFs will have an average kinetic energy per 
 nucleon of about 1 MeV and 0.5 MeV respectively, which is significantly lower than the average kinetic energy of the SNs, see Fig.~\ref{fig:ke}, given by 3.51 $\pm$ 0.25 MeV, 3.42 $\pm$ 0.27 MeV, and 2.67 $\pm$ 0.24 MeV for $^{236}$U, $^{240}$Pu, and $^{252}$Cf respectively.  As noted by R. Capote~\cite{Capote:2023}, our results, which are consistent with high-energy neutrons observed via dosimetry measurements~\cite{Schulc:2023}, point to an unmistakable need to include SNs in the analysis of prompt neutron spectra~\cite{Brown:2018}.
As a result, the FFs will never have a chance to catch up with them.  Additionally, the SNs are essentially free, since their total interaction energy, estimated using the 
neutron equation of state~\cite{Bulgac:2018},
\begin{equation}
  E_n^{\mathrm{int}} =\int dV[ a_n n_n^{5/3} + b_n n_n^2 + c_n n_n^{7/3}] \ll E_n^{\mathrm{kin}},
 \end{equation}
comprises less than 1 percent of their kinetic energy.  
The total number of emitted neutrons, shown in Fig.~\ref{fig:nsci}, is about $0.30\pm 0.05$, $0.26 \pm 0.05 $, and $0.55\pm0.02$ per fission event for $^{236}$U, $^{240}$Pu, 
 and $^{252}$Cf respectively, which is a considerable portion (roughly $9-14$ \%) of the total emitted prompt neutrons, see
 Refs.~\cite{Vandenbosch:1973,Wagemans:1991}. These are somewhat conservative estimates, see the  discussion in Ref.~\cite{supplement}, and these numbers can be likely enhanced by at least a factor of 1.25.  This is clear in Fig.~\ref{fig:nsci} where neither the emission of neutrons or protons has flattened. In comparison, Carjan \textit{et al.}~\cite{ Rizea:2008,Carjan:2010,Carjan:2015,Capote:2016,Carjan:2019} estimated an upper bound of $25 - 50$ \% of prompt fission neutrons are emitted during scission. At the same time the number of emitted protons is about 
 two orders of magnitude lower, see Fig.~\ref{fig:nsci}.  
The nucleons are emitted in roughly equal numbers both transverse to the scission axis and in front of the FFs.


  In summary, we have clarified several aspects of the most non-equilibrium and fastest stage of nuclear fission dynamics.  Within TDDFT, the neck 
rupture is not a random process, as previously argued in various phenomenological models.  Additionally, it appears that the neck rupture has similar  dynamics for a large class of asymmetric fission events, irrespective of nucleus considered or the initial conditions, beyond the top of the outer fission barrier.  This universality carries over to the emission of SNs, whose signal always appears as three distinct clouds, one transverse to the fission axis and two in front of each FF, in almost equal proportions.  
The aspects of the neck dynamics discussed above, 
can serve as a theoretical input for any semi-phenomenological approach to study FF properties~\cite{Becker:2013,Talou:2021,Vogt:2009,Litaize:2012}.

 The idea of SNs, proposed by \textcite{Bohr:1939}, is almost as old as nuclear fission itself.  The existence of SNs has been debated over the years~\cite{Wilson:1947, Debenedetti:1948, Stavinsky:1959, Fuller:1962, Bowman:1962, Kapoor:1963, Skarsvaag:1963, Gavron:1974, Boneh:1974, Pringle:1975,Franklyn:1978, Franklyn:1986, Jorgensen:1988,  Milek:1988,Samant:1995,Hwang:1999,Carjan:2007, Rizea:2008,Carjan:2010,Carjan:2015,Capote:2016b,Capote:2016,Guseva:2018, Vorobyev:2018, Carjan:2019}, see also \textit{Historical Note} in Ref.~\cite{Wagemans:1991}, and their experimental confirmation is still an open question. While neutron properties in earlier studies using simplified models~\cite{Rizea:2008,Capote:2016,Carjan:2019} have some features somewhat similar to what we find, they are missing the major component of emission perpendicular to the fission axis. The small fraction of SNs carry noticeable kinetic energy, and what is more surprising is that their energy spectrum ranges up to almost 18 MeV, as in shown in Fig.~\ref{fig:ke}, similar to other observations~\cite{Capote:2023,Brown:2018}. Along with SNs, a very small fraction of protons are emitted as well, and their fraction suggests a theoretical estimate for the emission of $\alpha$-particles 
and other charged nuclei.
   
   \vspace{0.5cm}         
{\bf Acknowledgements} \\

We thank Kyle Godbey and Guillaume Scamps for many useful discussions, and to Roberto Capote for his feedback regarding the impact of this work on the description of prompt fission neutron spectra.  The work of I.A. and I.S. was supported by the U.S. Department of Energy through the Los
Alamos National Laboratory. The Los Alamos National
Laboratory is operated by Triad National Security, LLC,
for the National Nuclear Security Administration of the U.S.
Department of Energy Contract No. 89233218CNA000001.
I.A. and I.S. gratefully acknowledge partial support and computational
resources provided by the Advanced Simulation and
Computing (ASC) Program.  The work of M.K. and A.B. was supported by the US DOE, Office of Science, Grant No. DE-FG02-97ER41014 and
also partially by NNSA cooperative Agreement
DE-NA0003841, and is greatly appreciated. 
This research used resources of the Oak Ridge
Leadership Computing Facility, which is a U.S. DOE Office of Science
User Facility supported under Contract No. DE-AC05-00OR22725. 


\providecommand{\selectlanguage}[1]{}
\renewcommand{\selectlanguage}[1]{}

\bibliography{local_fission}

\end{document}


\title{Supplement to Neck Rupture and Scission Neutrons in Nuclear Fission}


\author{{Ibrahim Abdurrahman}} 
\affiliation{Theoretical Division, Los Alamos National Laboratory, Los Alamos, NM 87545, USA}   
\author{{Matthew Kafker}}
\affiliation{Department of Physics, University of Washington, Seattle, WA 98195--1560, USA}
\author{{Aurel Bulgac}}
\affiliation{Department of Physics, University of Washington, Seattle, WA 98195--1560, USA}
\author{{Ionel Stetcu}}
\affiliation{Theoretical Division, Los Alamos National Laboratory, Los Alamos, NM 87545, USA}   


   
\date{\today}

\maketitle   

\begin{section}{Minimal history of scission neutrons}

\begin{figure*}
    \centering
    \includegraphics[width=0.99 \textwidth]{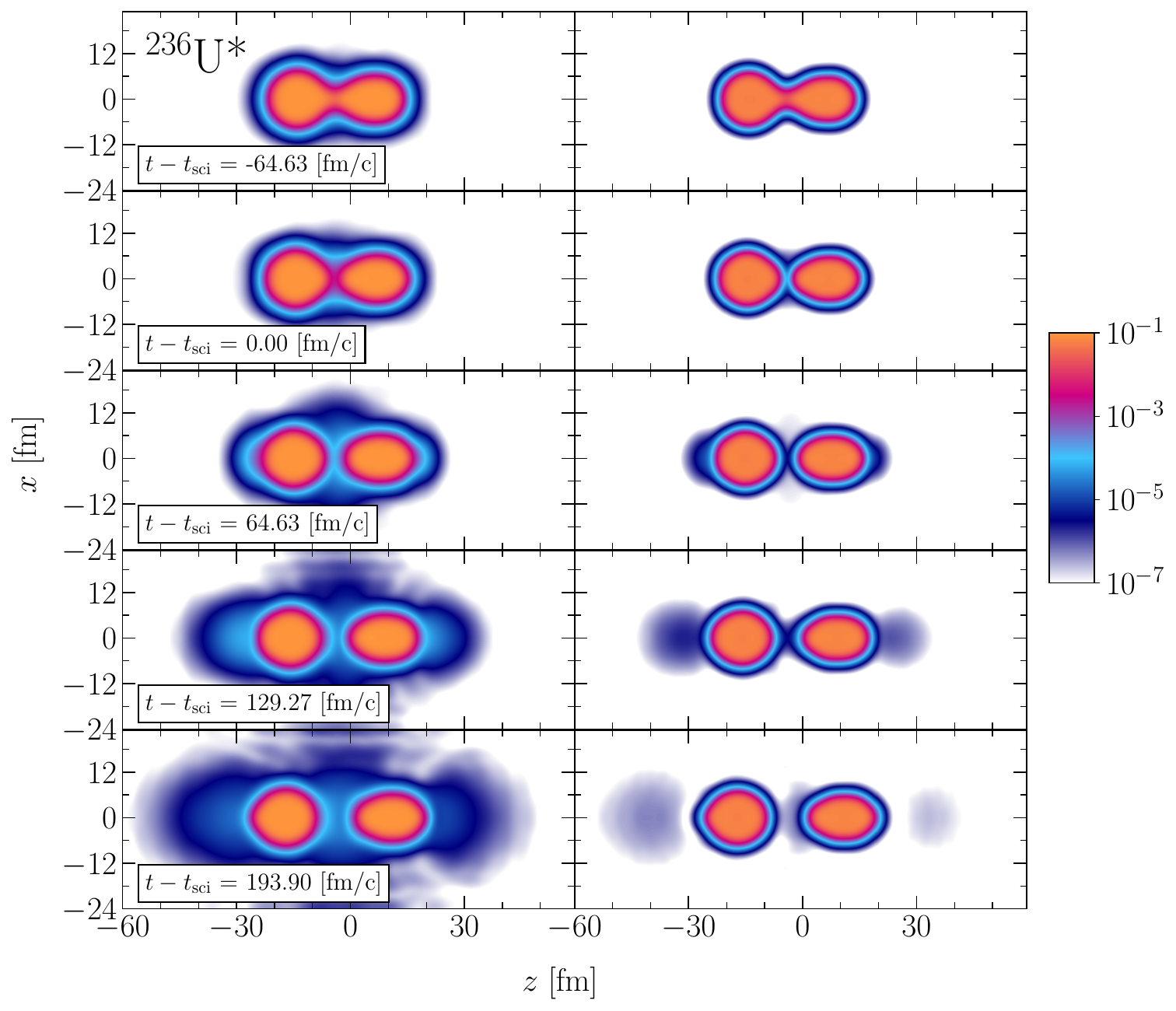}
    \caption{Left/right panels show time series of the neutron/proton number densities in fm$^{-3}$ for a typical fission trajectory.  }
    \label{fig:tseries}
\end{figure*}

\begin{figure}
    \centering
    \includegraphics[width=0.5 \textwidth]{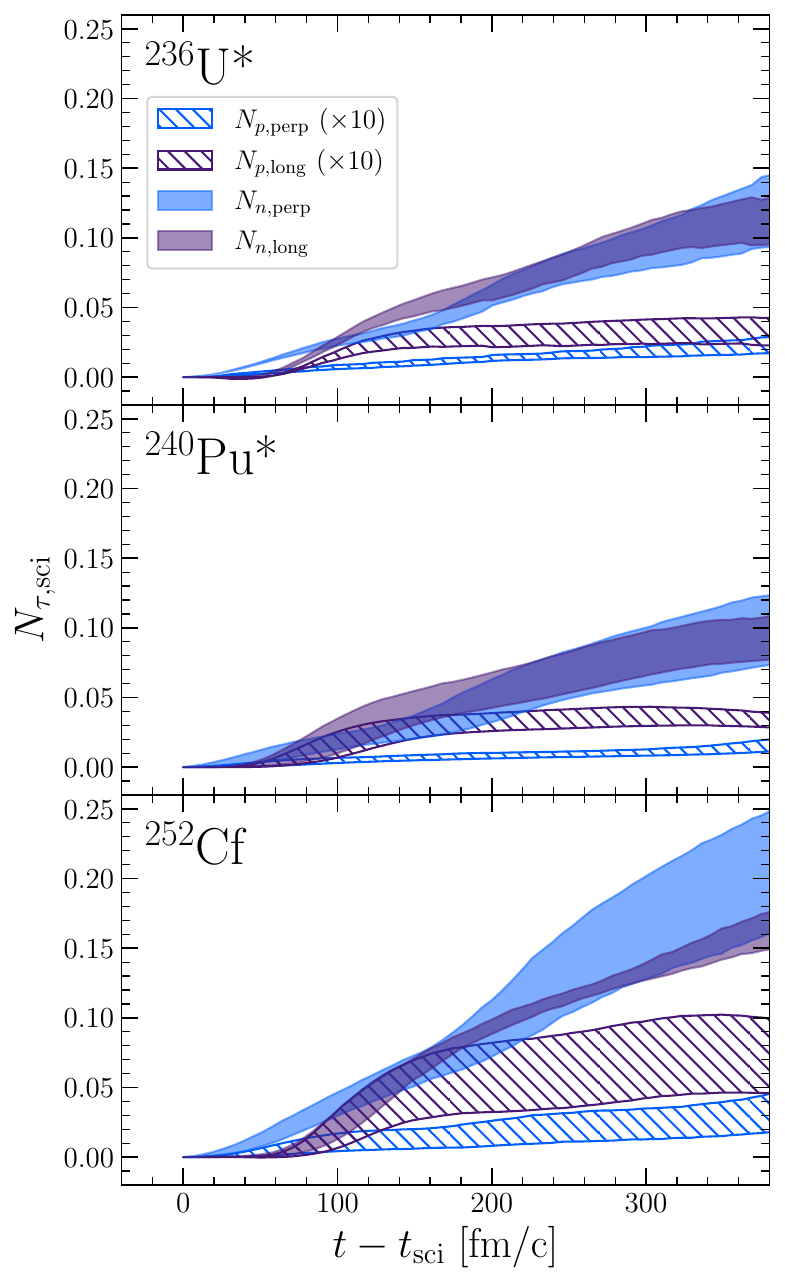}
    \caption{The number of scission protons (dashed lines) and scission neutrons (solid lines) are shown as functions of time for various nuclei up to when the longitudinal clouds hit the boundaries of the box.  The perpendicular and longitudinal components (with respect to the fission axis) are shown separately.  The scission proton component has been scaled by a factor of 10. The number of scission neutrons in the perpendicular and longitudinal directions are roughly equal, meanwhile scission protons favor being emitted in front of the FFs.}
    \label{fig:dirs}
\end{figure}

\begin{figure*}
    \centering
    \includegraphics[width=0.99 \textwidth]{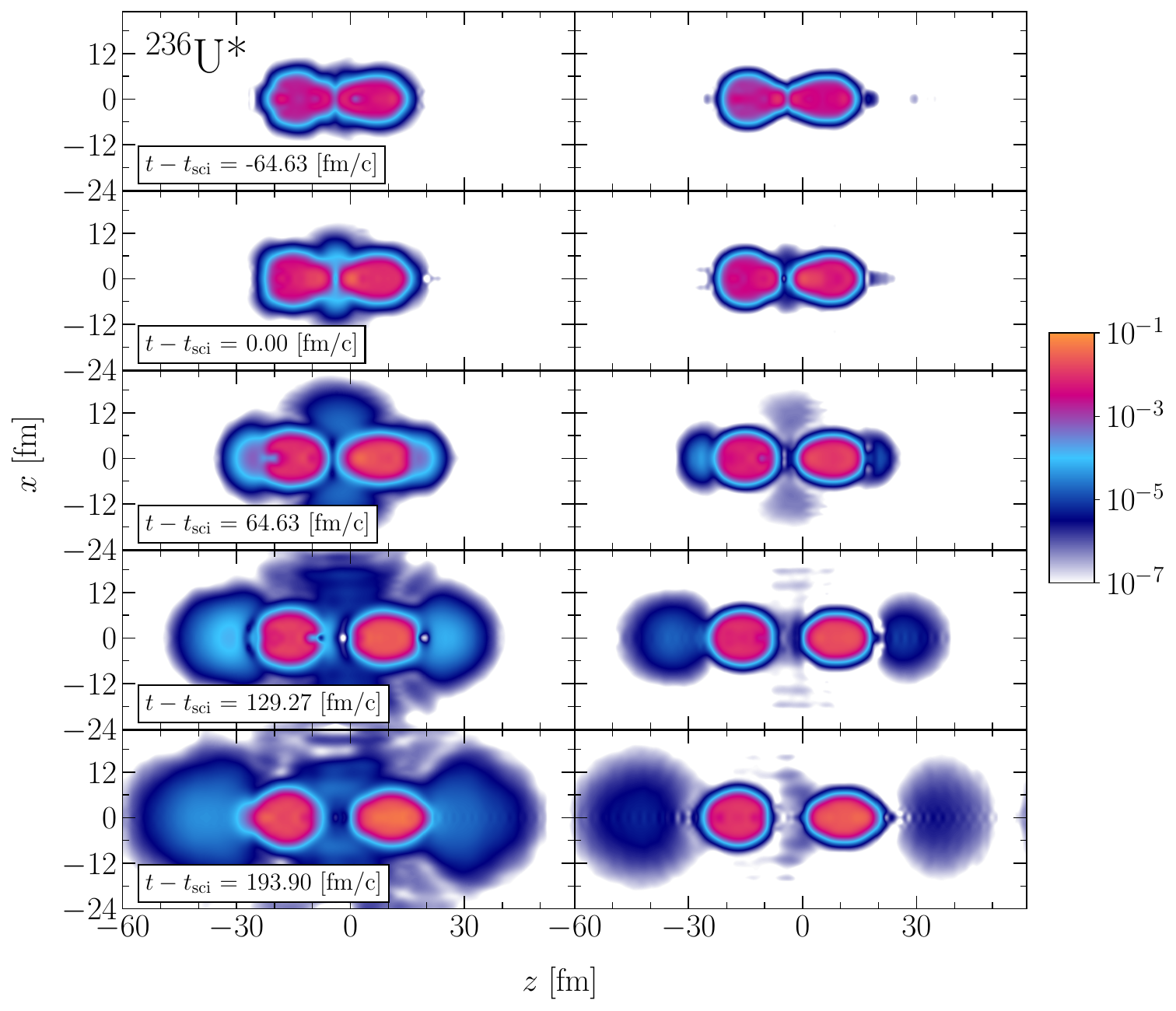} 
    \caption{Left/right panels show time series of the neutron/proton collective flow energy densities, $ \mathcal{E}_{\tau,\mathrm{coll}}  $, in MeV/fm$^3$ for a typical fission trajectory.  }
    \label{fig:eflow}
\end{figure*}

\begin{figure*}
    \centering
    \includegraphics[width=0.99 \textwidth]{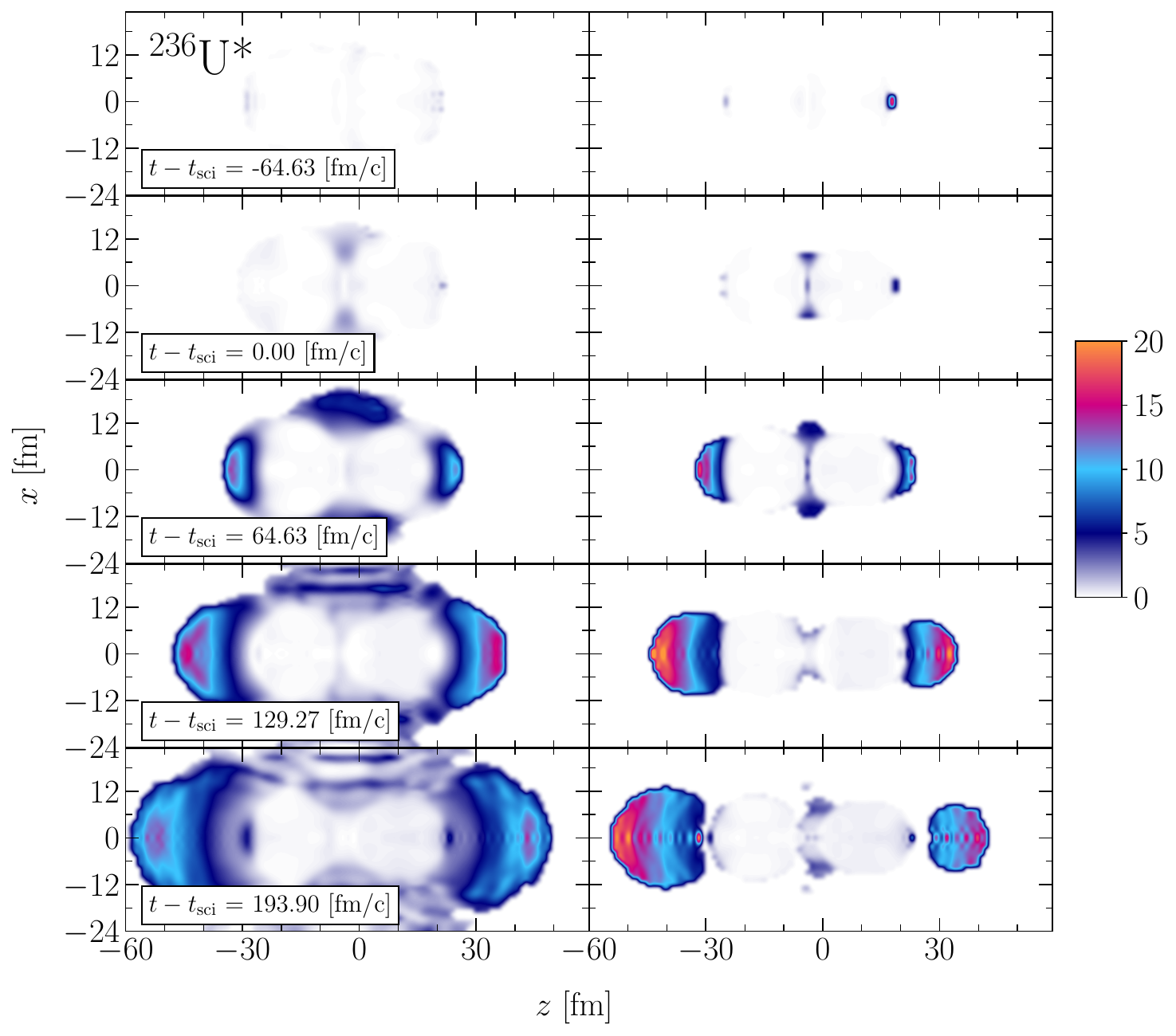}
    \caption{Left/right panels show time series of the neutron/proton collective flow energy per nucleon in each cell, $ \frac{\mathcal{E}_{\tau,\mathrm{coll}}}{n_\tau} $, (values given in the colorbar) for a typical fission trajectory in units of MeV. Cells with $n_\tau < 10^{-7}$ were excluded. Within the bulk of the FFs, defined by $n_n > 10^{-2}$, $ \frac{\mathcal{E}_{n,\mathrm{coll}}}{n_n} $ varies between $\sim$ 0.1 to 0.5 MeV. }
    \label{fig:eflow2}
\end{figure*}

\begin{figure}
    \centering
    \includegraphics[width=0.5 \textwidth]{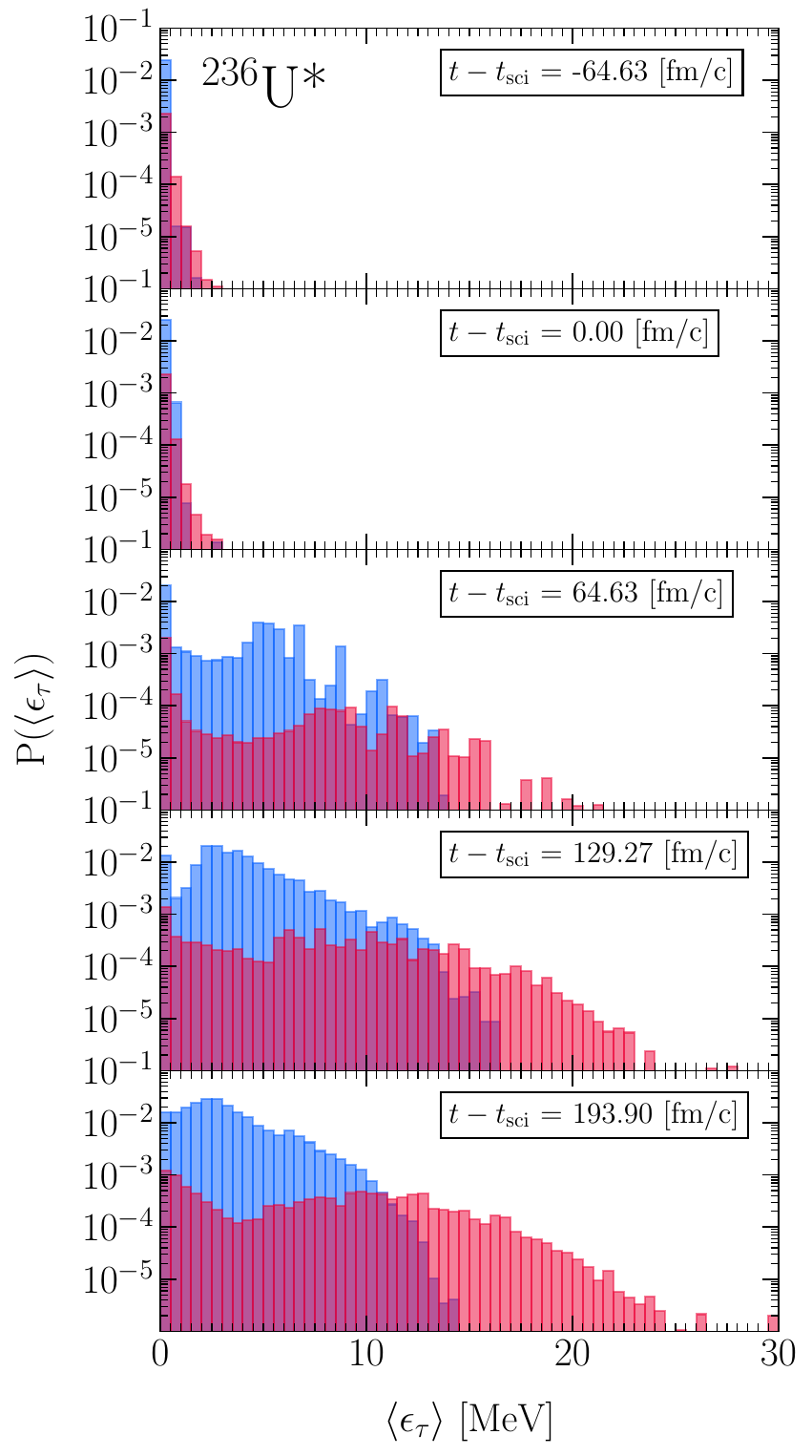}
    \caption{Time series of scission neutron/proton (blue/red) collective flow energy distributions for a typical fission trajectory.  The distributions are normalized in time to the number of scission neutrons/protons: $\sum P(t,\langle \epsilon_\tau \rangle) \Delta \epsilon= N(t)_{\tau, \mathrm{sci}}$, with $\tau=n,p$.  The distribution was obtained by computing the collective kinetic energy per nucleon at each cell on the lattice, $\langle \epsilon_\tau \rangle = \frac{1}{2} m v_\tau^2$, and placing the point in the appropriate bin. The contribution of each point is weighted by the number density within the same cell. } 
    \label{fig:eflowdis}
\end{figure}

\begin{figure}
    \centering
   \includegraphics[width=1 \columnwidth]{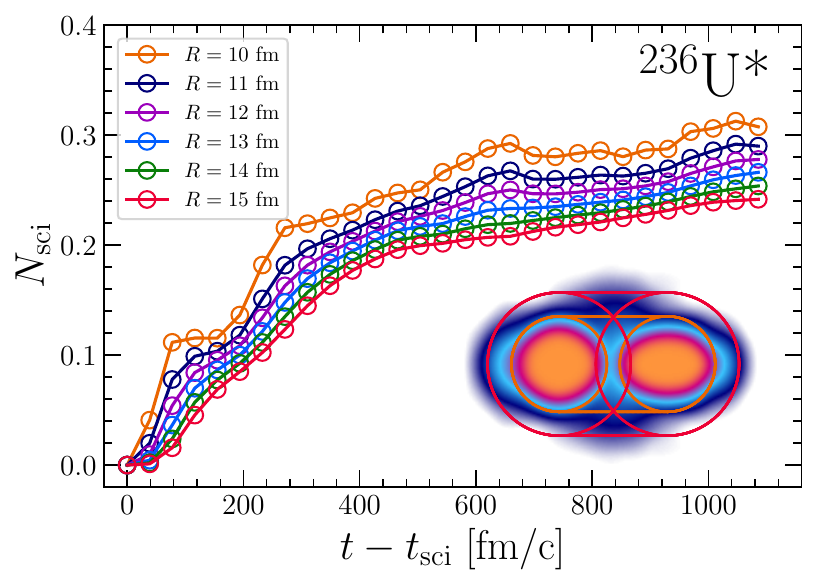}
    \includegraphics[width=1 \columnwidth]{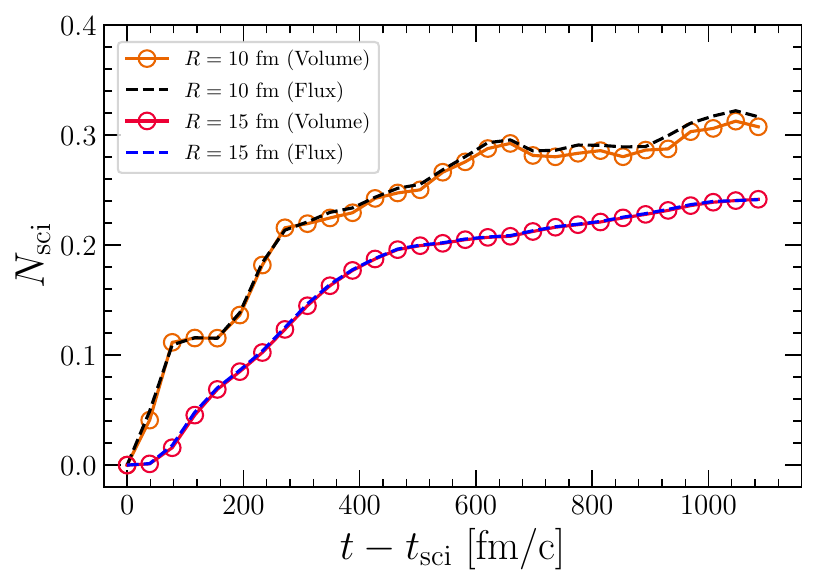}
    \includegraphics[width=1 \columnwidth]{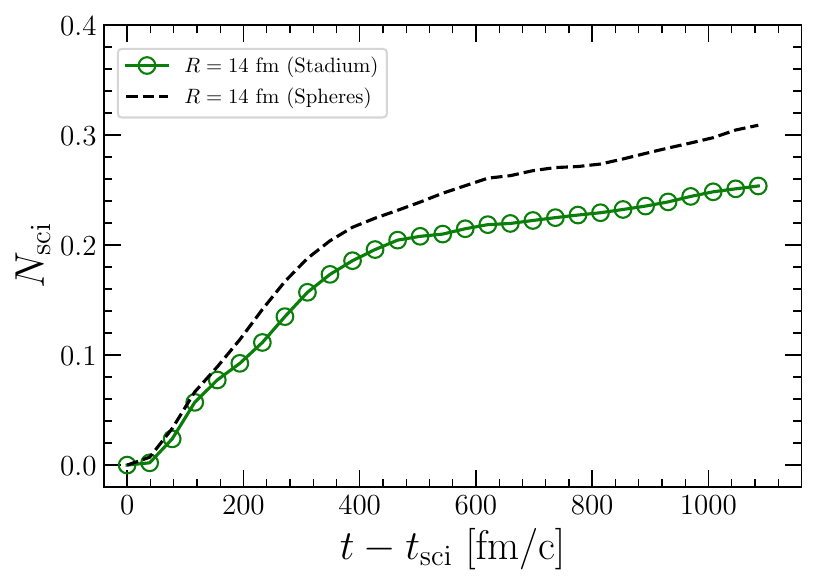}
    \caption{ (Top panel) the number of scission neutrons obtained by integrating the volume outside of bounding stadiums with different radii (see inset).  The maximum relative percent difference between the stadiums with $R = 10$ fm and $R = 15$ fm is $\sim$25\%. (Middle panel) comparison between computing the scission neutrons via $N_{\mathrm{sci}} = \int_{\mathrm{Vol}} d\bm{r}n_n(\bm{r}) $ and $N_{\mathrm{sci}} = \int_{\mathrm{Surf}} d S \bm{\nabla} \cdot \bm{j} (\bm{r}) $, where the volume refers to the region of space outside of the stadium. The maximum relative percent difference these two methods is given by $\sim$5\%.  (Bottom panel) comparison between two bounding containers (see inset of top panel) with $R = 14$ fm. The maximum relative percent difference is given by $\sim20\%$. The bounding spheres are what was used for all other results in this study.   
    }
    \label{fig:comp} 
    \end{figure}

Nuclear fission was experimentally discovered by \textcite{Hahn:1939} in 1939.  In the same year, it was named and its main mechanism was explained by \textcite{Meitner:1939}. Surprisingly, the idea of scission neutrons (SNs) is almost as old as the discovery of fission itself, also being first considered in 1939.  At the time, Bohr and Wheeler conjectured three sources for neutron emission during fission: first delayed neutrons occurring on the time scale of seconds, second neutrons evaporated from the fission fragments (FFs) as a result of their excitation, now known as prompt neutrons, and finally, and much more speculatively, neutrons formed due to the neck rupture~\cite{Bohr:1939}:\\ \\
\emph{We consider briefly the third possibility that the neutrons in question are produced during the fission process itself.  In this connection attention may be called to observations on the manner in which a fluid mass of unstable form divides into two smaller masses of greater stability; it is found that tiny droplets are generally formed in the space where the original enveloping surface was torn apart.}
\\ \\ 
For a long time after, the idea remained dormant.  In the late 40s, the first experiments investigating the directional properties of neutrons, produced from the fission of $^{235}$U(n,f), were conducted.  At the time there was a consensus that the experimental results were consistent with the assumption that all neutrons are isotropically emitted from fully accelerated FFs, and any brief consideration for SNs were quickly dismissed~\cite{Debenedetti:1948}. As experiments were refined in the 50s conclusions remained the same~\cite{Fraser:1952,Fraser:1954}.  It was not until the 60s, starting with experiments conducted by \textcite{Bowman:1962}, that the idea of SNs would re-emerge with considerable momentum. 

In 1962, this group computed the angular distributions of neutrons emitted from the spontaneous fission of $^{252}$Cf, observing deviations from the "isotropic hypothesis" (which states all neutrons would be emitted isotropically in the intrinsic frame of fully accelerated FFs).  Using a model for SNs proposed by \textcite{Stavinsky:1959} in 1959 and \textcite{Fuller:1962} in 1962, who extended the work of \textcite{Halpern:1959} from scission alphas to scission neutrons, \textcite{Bowman:1962} predicted that SNs would comprise $\sim$ 10\% of prompt neutrons emitted during fission. Such models assume SNs are treated as being isotropically emitted from the region of the neck rupture. In particular, Fuller proposed the sudden rise of the nuclear potential at the neck during scission would lead to an expulsion of nuclear matter, and estimated scission neutrons would comprise 10-20\% of neutrons emitted, in the case of the spontaneous fission $^{252}$Cf, with each carrying an average of 2-6 MeV of energy~\cite{Fuller:1962}.  Subsequent experiments mostly agreed with the findings of \textcite{Bowman:1962,Kapoor:1963,Skarsvaag:1963}, with one proposing an alternative explanation to account for inconsistencies between experimental results and the isotropic emission hypothesis, namely pre-scission neutron emission after the nucleus crosses the saddle~\cite{Kapoor:1963}. 

It is important to re-stress, conclusions pertaining to the neck rupture and related phenomena, such as SNs, cannot be probed directly by experiment.  As a result, all specific conclusions about the emission mechanism of SNs are extremely model dependent, and, beyond the 60s, experimental efforts to confirm the existence of and/or estimate the effects of SNs led to mixed results, with estimates placing the number of SNs as low as 1\% to as high as 15\%~\cite{Gavron:1974, Pringle:1975, Franklyn:1978, Franklyn:1986, Jorgensen:1988, Samant:1995, Hwang:1999, Vorobyev:2010, Guseva:2018, Vorobyev:2018}.  Significantly, the vast majority assumed a simple model for SN emission, namely, all SNs are expelled isotropically in the laboratory frame, close to the neck rupture, and typically with low kinetic energy.  This, despite the greater variance in  the theoretical models available over time.  To illustrate the great historical confusion surrounding the topic consider the following quote from~\textcite{Wagemans:1991} in 1991: \\ \\
\emph{Up to a few years ago, it was generally admitted that 15 to 20\% of neutrons emitted during the fission process were scission neutrons.  This has been contradicted by recent results ...  These measurements indicated that probably only 1.1 $\pm$ 0.3\% of the neutrons are scission neutrons and that their productions mechanism is similar to that of satellite droplets in the disintegration of liquid jets.}
\\ 

In general, the majority of models proposed and/or used to model SN emission can be classified into three distinct classes. First, neutrons emitted due to the sudden change in the nuclear potential during the neck rupture, historically, the most commonly evoked model~\cite{Boneh:1974, Carjan:2007, Milek:1988}. In this case the number and kinetic energies of the SNs are determined by the amount of matter in the neck region and the time it takes the neck to rupture. The majority of SNs are expected to have higher kinetic energy (relative to prompt neutrons), and either be emitted isotropically from the region of the neck rupture or (in some models) perpendicular to the scission axis.  Second, fragments formed as satellite droplets~\cite{Brosa:1992}, which could comprise both neutrons and protons, in cases of fission where the neck is highly elongated~\cite{Brosa:1990}.  This mechanism would have a similar signal to the first case, except the kinetic energies of the SNs would be significantly lower.  Last, SNs emitted in front the FFs via the so called catapult mechanism: as the neck ruptures the tails of the densities of FFs are "snatched" inside, travel through the nucleus with high energy, and emerge on the other-side as SNs~\cite{Madler:1985}. Here, the majority of the SN signal would be seen as polar emission (in the direction of the FFs) with high kinetic energy, all with respect to the lab frame.  All of the above proposed mechanisms have distinct features, and hence unique signals for determining which is primarily responsible for SNs. In addition, recent hybrid approaches, based on the Time-Dependent Schr$\rm \ddot{o}$dinger Equation in polar coordinates (TDSE2D), have been used to model SNs~\cite{Rizea:2008, Carjan:2010, Carjan:2015, Capote:2016, Carjan:2019}.  In some microscopic studies, pairing correlations were treated in the BCS approximation in two different limits, using frozen occupation numbers or allowing them to change via the BCS equations~\cite{Rizea:2008,Capote:2016,Carjan:2019}.  Frozen occupation numbers are unrealistic, as we have shown the population of levels change in time for nuclear systems \cite{Stetcu:2011}. Allowing states to change their occupation numbers in TDHF+TDBCS theory, see e.g. Ref.~\cite{Tanimura:2017,Ren:2022}, leads to the violation of the continuity equation~\cite{Scamps:2012}, which is a crucial ingredient in describing matter transport at both the classical and quantum level.  In the most recent study of the neck formation~\cite{Ren:2022}, which has no mention of SNs however, the authors relied on the use of the nucleon localization function (NLF), a concept we find highly debatable~\cite{Bulgac:2023}. The authors make the claim that the neck ruptures because of the Coulomb repulsion of the two proton pairs, identified in the neck region using the NLF. One can easily evaluate the Coulomb force between such two proton pairs to be about 0.35 MeV/fm, which is almost two orders of magnitude smaller than the Coulomb repulsion between two touching FFs of ${\cal O}(10)$ MeV/fm, which is the real cause of the separation of the two FFs. As is well known, the Coulomb repulsion between two touching  uniformly charged spheres is equal to the Coulomb repulsion of two point charges with charge equal to the FFs' charges separated by the same distance.   

The Brosa model, in which SNs were never discussed, {\protect\cite{Brosa:1990}} introduced concepts in fission such as standard, superlong, supershort, and superasymmetric pre-scission shapes, which some practitioners find useful.  Some of these modes have not been observed in microscopic models.  This suggests such shapes have a very low probability to occur.

\end{section}

\begin{section}{ Charged Particle Emission during the Neck Rupture}

Unlike in the case of SNs, it has been well known since the late 40s that alpha particles are emitted at scission~\cite{Demers:1946,Farwell:1947,Wollan:1947,Marshall:1949,Titterton:1951,Allen:1950,Feather:1947,Nobles:1962}.  Coined long range alphas (LRAs), the vast majority emerge perpendicular to the fission axis, with a much smaller fraction alphas with energies greater than 25 MeV are emitted in the direction of the FFs~\cite{Fraenkel:1967}. Because most LRAs are emitted perpendicular to fission axis, the majority must originate from the neck during rupture. Later, additional charged ternary particles were detected, including Tritium~\cite{Albenesius:1959,Watson:1961,Marshall:1966}, $^6$He~\cite{Goward:1949,Whetstone:1965,Dakowski:1967,Cosper:1967,Krogulski:1969,Chwaszczewska:1967}, isotopes of Li and Be~\cite{Vorobiev:1969}, and heavier elements~\cite{Vandenbosch:1973,Rose:1984,Poenaru:1986}.  

Our TDDFT simulations show proton clouds emitted in front of the FFs, with significantly lower probability than the neutron clouds, see Figs.~\ref{fig:tseries}, \ref{fig:dirs}. Additionally, scission protons favor emission in the direction of the FFs instead of perpendicular to the FF axis, unlike scission neutrons, which will emit in both directions with roughly equal probability. These estimates of scission proton probabilities can act as a theoretical estimate for the emission of light charged nuclei.

Current TDDFT simulations cannot describe the potential condensation of light charged particles within this cloud.  As it is well known from condensation studies, e.g. in the last three decades of cold atom studies~\cite{Dalfovo:1999,Pethick:2008,Giorgini:2008}, the condensation, which determines the lifetimes of these systems, is controlled by the very low rate of three-body collisions, as only when three particles collide can a two-body bound state form, which is the first step towards condensation. Similar arguments are used in the formation of hadrons from quarks in relativistic heavy-ion collisions, or the formation of rain drops in the fog. The probability to find three nucleons in close proximity to each other is extremely low in the case of scission neutrons, and furthermore, only two-body collisions are included in TDDFT so far. Even in quantum kinetic models, far less computationally demanding theories, the inclusion of three-body collisions is not a simple procedure.  For example, the most common collision integrals only contain two body collisions~\cite{Fowler:1928,Uehling:1933,Bertsch:1988}.  

An alternative and distinct mechanism to condensation for the formation of alphas, deuterium, tritium, and other heavier light nuclei is ternary fission of preformed fragments, which was observed since the 1960’s, see~\textcite{Vandenbosch:1973} and~\textcite{Rose:1984}.  In principle this mechanism is present in TDDFT, but in practice it would require 2 to 4 orders of magnitude more fission events to be simulated, depending on the mass of the charged nuclei, which is not computationally feasible at this time.  One expects that light nuclei are formed in the neck region, and not in the clouds of emitted nuclear matter, although experimentally one could not distinguish between the two types unless very different angular distributions are predicted and observed.  The present nuclear formulation of TDDFT does not include proton-neutron correlations and three and higher body correlations, which can affect these processes.  In Ref.~\cite{Ren:2022} the authors claim to have seen the formation of two alpha particles in the neck region for the induced fission of $^{240}$Pu using TDDFT with the BCS approximation.  This process has not been validated experimentally and we dispute the claim in Ref.~\cite{Bulgac:2023b}, as in such simulations the alpha particles never actually escape the compound system or FFs.

\end{section}

\begin{section}{Collective Flow Energy}

The collective flow energy of the scission particles is given by,
\begin{equation}
E_{\mathrm{coll}} = \int_{V_{\mathrm{sci}}} \mathcal{E}_{\mathrm{coll}} (\bm{r}) d^3 \bm{r} = \int_{V_{\mathrm{sci}}} \frac{1}{2} m n_\tau v_\tau^2 d^3 \bm{r},
\end{equation}
where $\tau = n$ or $p$; $V_{\mathrm{sci}}$ refers to the volume containing the scission particles, excluding the FFs; m refers to the mass of a nucleon, taken as the average of the neutron and proton rest masses, and $v_\tau = \frac{\hbar j_\tau}{m n_\tau}$.  For additional information about the collective flow energy please see Refs.~\cite{Bulgac:2016,Bulgac:2019c,Bulgac:2020}.

In Figs.~\ref{fig:eflow}, \ref{fig:eflow2} the collective flow energy density, 
\begin{equation}
\mathcal{E}_{\tau,\mathrm{coll}} = \frac{1}{2} m n_\tau v_\tau^2 = \frac{\hbar^2 j_\tau^2}{2 m n_\tau }
\end{equation}
and collective flow energy per nucleon within each cell,
\begin{equation}
\frac{\mathcal{E}_{\tau,\mathrm{coll}}}{n_\tau} = \frac{1}{2} m v_\tau^2 = \frac{\hbar^2 j_\tau^2}{2 m n_\tau^2 }
\end{equation}
are shown for a single fission trajectory.  The majority of the scission particles contain a larger flow energy per nucleon than the FFs during the neck rupture.  Also, a small subset contain a much larger collective flow energy, up to $\sim$ 15 MeV for neutrons and $\sim$ 20 MeV for protons, relative to $\sim$ 0.1 to 0.5 MeV for the bulk of nucleons in the FFs.  

Fig.~\ref{fig:eflowdis} shows the distribution of the collective flow energy per nucleon for scission neutrons and protons. 
The collective flow energy can not be tracked for long times after the neck rupture, due to current limitations in the lattice size.  More specifically, once scission particles reach the other side of box they begin to collide with images of FFs, due to the periodic boundary conditions of the lattice, causing the flow energy to unphysically decrease.

In the main manuscript we focused on the kinetic energy, 
\begin{align}
E_{\mathrm{kin}} = \int \frac{\hbar^2 \tau}{2 m} d^3 \bm{r}, 
\end{align}
as opposed to the collective flow energy, since the former is less sensitive to the size of the lattice.  The kinetic energy accounts for both the collective and intrinsic motion of the nucleons, while the collective flow energy only accounts for the former and hence vanishes for stationary states.  Both have similar values for scission particles until the clouds reach the boundaries, in which case the kinetic energy continues to increase and the collective energy decreases.

The kinetic energy distribution in the main manuscript was obtained by recording the kinetic energy per scission neutron, $\frac{E_{\mathrm{kin}}}{ N_{\mathrm{sci}}}$, over time.  The distribution was generated by points defined via:  
\begin{equation}
E_{n,\mathrm{kin}} (t) = E_{n,\mathrm{kin}}(t) - E_{n,\mathrm{kin}}(t-\Delta t) ,
\end{equation}
\begin{equation}
N_{n,\mathrm{sci}}(t) = N_{n,\mathrm{sci}}(t) - N_{n,\mathrm{sci}}(t-\Delta t),
\end{equation}
weighted by $N_{n,\mathrm{sci}}(t) - N_{n,\mathrm{sci}}(t-\Delta t)$ in order avoid double counting.  The extraction of the distribution of the collective flow energy per nucleon is explained in Fig.~\ref{fig:eflowdis}. Numerically, we cannot calculate the kinetic energy per nucleon in each cell, $ \frac{\hbar^2 \tau_\tau}{2 m n_\tau}$. 

\end{section}
\begin{section}{Methods}\label{sec:meth}

\begin{table*}
\begin{tabular}{  c c c c c c  } 
  \hline
    \hline
  Run \# & Nucleus & $Q_{20}$ [b]& $Q_{30}$ [b$^{3/2}$]& \quad $\beta_{2}$ \quad & \quad $\beta_{3}$ \quad \\ 
  \hline
  1 & $^{236}$U & 184.33	& 19.66 & 1.88 & 0.86 \\
  2 & $^{236}$U & 159.64	& 17.80 & 1.63 & 0.77 \\
  3 & $^{236}$U & 135.25	& 12.74 & 1.38 & 0.55 \\
  \hline
  1 & $^{240}$Pu & 157.20	& 20.18	& 1.56 & 0.85 \\
  2 & $^{240}$Pu & 153.11	& 18.34	& 1.52 & 0.77 \\
  3 & $^{240}$Pu & 140.08	& 10.6	& 1.39 & 0.45 \\
  4 & $^{240}$Pu & 141.85	& 8.56	& 1.40 & 0.36 \\
  5 & $^{240}$Pu & 144.71	& 6.63	& 1.43 & 0.28 \\
  6 & $^{240}$Pu & 145.64	& 6.63	& 1.44 & 0.28 \\ 
  \hline
  1 & $^{252}$Cf & 240.80	& 36.53	& 2.20 & 1.39 \\
  2 & $^{252}$Cf & 227.19	& 32.50	& 2.07 & 1.24 \\
  3 & $^{252}$Cf & 199.17	& 23.52	& 1.82 & 0.90 \\
  4 & $^{252}$Cf & 168.29	& 13.44	& 1.54 & 0.51 \\
  \hline
    \hline
  \end{tabular}
    \caption{Initial deformation parameters of fission trajectories. For $^{240}$Pu initial conditions were chosen both above and below the saddle point (for more details see~\cite{Bulgac:2019c,Bulgac:2020}). }
    \label{tab:ini}
\end{table*}

All TDDFT simulations were performed with the LISE package, a solver for both the static and time-dependent super-fluid local density approximation (SLDA and TDSLDA respectively) equations in three dimensions~\cite{Shi:2020}.
(TD)SLDA is a general framework, introduced, tested, and verified not only in case of nuclei, but also in case of  cold atoms and neutron stars, both in equilibrium and out of equilibrium, and confronted with ab initio quantum results for many-body systems and which led in a number of cases to a correct interpretation of experimental data, see Ref.~\cite{Bulgac:2019} for a relatively recent review. (TD)SLDA is free of uncontrolled approximations
and unchecked assumptions, and relies on a very small number of well established properties of the many-body system in question. In case of nuclear systems these are the binding energy and the saturation density of symmetric nuclear matter, surface tension, proton charge, symmetry energy and to a lesser degree its density dependence, and the strengths of the spin-orbit and pairing interactions, with which one obtains one of the most accurate mean field predictions of nuclear masses, charge radii, shell structure, one and two-nucleon separation energies, etc.~\cite{Bulgac:2018}.

Conclusions drawn in the manuscript should hold if fluctuations are included and if symmetries, corresponding to proper quantum numbers, are restored.  In the language of TDDFT, both refer to mixing different trajectories or orientations.  We include vastly different trajectories (deformations, excitation energies, fission nuclei) and observe the same behavior, thus it is natural to expect the mixing of trajectories is not going to change our conclusions, at least for one-body observables, which is all we consider so far. In the language of Generator Coordinate
Method, these configurations provide an almost complete
basis set of states to describe the nucleus right after it passes the second well. Furthermore, mixing trajectories cannot lead to dramatic changes on a time scale as short as the neck rupture.  That said, in the saddle-to-scission segment of the fission dynamics, quantum fluctuations will likely be important, which is why we consider many different points along the outer saddle.   

All initial conditions were chosen beyond the outer saddle, with energies equal to the outer saddle for compound systems $^{236}$U and $^{240}$Pu, and energies equal to the ground state for $^{252}$Cf.  A total of 3/6/4 trajectories were ran for $^{236}$U/$^{240}$Pu/$^{252}$Cf respectively.  The deformation parameters for all initial conditions are given in Table.~\ref{tab:ini}.  $^{236}$U calculations were performed in a $N_x=48$, $N_y=48$, $N_z=120$ lattice with spacing 1 fm on OLCF's supercomputer Summit.  $^{240}$Pu and $^{252}$Cf calculations were performed in a $N_x=48$, $N_y=48$, $N_z=100$ lattice with spacing 1 fm on Lawrence Livermore National Laboratory's supercomputer Sierra.

Of importance to highlight is the enormous computational effort required to run even a single trajectory.  As an example consider one $^{236}$U fission simulation on Summit. A total of 1,105,920 qpwfs, each with four spatial components,  were evolved for $\sim$40,000 time-steps, equivalent to solving 4,423,680 nonlinear coupled partial differential equations in 3D+time for the same number of time-steps. $4,423,680 = 2\times 2\times 4\times 48^2\times120$, where $4\times 48^2\times120$ is for the dimension of the HFB Hamiltonian, $2\times 4\times 48^2\times120$ is the number of either proton or neutron quasiparticle eigenstates, and the first factor of 2 is accounting for the presence of both protons and neutrons respectively.  Each quasiparticle wave function has one $u$ and one $v$ components respectively, each of them with two spin substates, and each spin component depending on $48^2\times120$ spatial coordinates. During the time evolution the entire set of quasiparticle states $2\times 4\times 48^2\times120$ participates in the dynamics, and using a restricted set of quasiparticle wave functions or a time-dependent Bardeen-Cooper-Shrieffer approximation, which violates the continuity equation~\cite{Scamps:2012}, leads often to very large errors or even numerically unstable time-evolutions~\cite{Bulgac:2023b}.
In our case, the evolution is extremely stable with energy conserved at the level of 100 keV or $\Delta E/E(t=0) < 5 \times 10^{-5} $ and particle number conserved at the level of $\Delta N/N(t=0) < 10^{-9} $. 

At each time-step roughly 213 TB of unique data is created, and even more is stored. The evolution of the system from the outer saddle until the FFs center of masses are separated by 90 fm required 4,608 nodes, corresponding to 27,648 Nvidia Tesla V100 GPUs, for $\sim$14 hours on Summit.  Each GPU contained 16 GB of high bandwidth memory (HBM2) and a theoretical double-precision flop rate of 40 TFLOPs.  Cumulatively, this is equivalent to 442 TBs of memory on the GPUs.  At the time such calculations were performed, and potentially to this day, it was likely the largest single computational simulation ran for such a long continuous wall-time. Similar, but slightly smaller calculations in boxes $48^2\times 90$ were performed using roughly 4,150 nodes, corresponding to 16,600 Nvidia Tesla V100 GPUs, on the Sierra supercomputer, each taking about 6 hours wall time. Despite using all of Summit and Sierra, we cannot track the SNs for too long before they hit the boundary.  To avoid this issue, we will need to increase the lattice by at least a factor of 2 in all spatial dimensions, corresponding to a total memory increase of a factor of 64. Even Frontier, the leading supercomputer as of 2024, does not offer the capability to perform such a simulation.  

In addition, a number of approaches were considered to compute the number of SNs.  Even though we considered various radii of these spheres, we settled on conservatively wrapping the two FFs with spheres with radii $R=14$ fm centered around the FFs center of masses, and integrating nuclear density outside of the spheres. One can make the case however that radii $r=10$ fm are also acceptable, in which case the number of scission neutrons increases by about 25\%.  We also considered a stadium, changing the radii of the bounding surface, and computing the flux passing through the surface.  We found minimal differences between the varying approaching, as shown in Fig.~\ref{fig:comp}. The number of  SNs was estimated in two ways, either by integrating their density outside the spheres/stadium or by evaluating the flux through these surfaces. The small differences between the two methods illustrate the small  relevance of re-absorption of the nucleons. 

\end{section}

\begin{section}{On the use of a Potential Energy Surface for Fission Calculations}

In molecular physics, the concept of the potential energy surface (PES) was developed in the Born-Oppenheimer approximation in 1927 and the role of non-adiabatic transitions plays a crucial role in many chemical reactions, see~\cite{Tully:1971,Tully:1990,Hammes:1994}. During the shape evolution of molecular systems  the energy levels for fixed shapes cross and the system jumps from one PES to another. 

The concept of a PES for a nucleus undergoing deformation was first introduced by~\textcite{Bohr:1939}. It is a natural evolution of Bohr’s semi-classical hydrogen atom model of 1911, where the 1s orbital (and all other orbitals) are treated as quantized circles.  The model predicted the correct ground state energy, but contained the wrong physics, as the real ground state of the hydrogen atom is a fuzzy sphere of about the same radius. This highlights the importance of a having a fully quantum description. In analogous fashion, it is now widely recognized  that the validity of the concept of a PES outside the outer saddle is highly questionable~\cite{Bender:2020}.  In the region outside the outer fission barrier the emerging fission fragments are getting hotter and hotter on the way to scission. The force between them depends on their instantaneous temperature, and is not determined by the gradient of the lowest PES with respect to the change in shape. Outside of the outer saddle, fission dynamics is controlled mostly by the gradient of the local entropy of the system, see~\cite{Frobrich:1998,Bulgac:2019c,Bulgac:2020}.  As shown in~\cite{Bulgac:2019c}, the gradient of the intrinsic energy (which can be interpreted as the intrinsic free energy) is significantly smaller than the gradient of the lowest PES, indicative of strongly damped dynamics. 

The damped dynamics were first demonstrated theoretically within a microscopic framework in~\cite{Bulgac:2016} and subsequent papers~\cite{Bulgac:2019c,Bulgac:2020}.  In~\cite{Bulgac:2016}, for the first time in the literature, real-time fission dynamics were studied starting from the top of the outer saddle until the FF fully separate. The Langevin/Fokker-Plank treatment of fission is only valid in the case of weak dissipation, where the use of a (standard) PES \`a la Bohr and Wheeler is reasonable.  This is not the case here, as the nuclear level density becomes exponentially large beyond the saddle, and the force acting generated by the collective degrees of freedom is basically vanishing in comparison.  Currently, only the phenomenological model of~\textcite{Albertsson:2020} is consistent with ~\cite{Bulgac:2016,Bulgac:2019c,Bulgac:2020}, where the authors use entropy-based dynamics instead of a Langevin approach. 

The fact that the PES is a useful concept in phenomenological models is the result of the relation between the PES and the nuclear level density at the excitation energy where fission occurs, instead of the result of the collective effects from the nuclear shapes on the surface. The nuclear level density, which determines the entropy, is the key driving factor in fission.  At the saddle configuration the nuclear level density is very low, comparable to the level density of a nucleus near the ground state. This argument goes back A. Bohr in 1956: see discussion at the start of Ch. V. in~\textcite{Vandenbosch:1973}.  Even though the compound nucleus evolves from inside the second fission well, where the level density is relatively high in case of neutron induced fission, once it arrives at the saddle, the nucleus is in a configuration very similar to the minimum energy solution at the saddle point, where the local nuclear level density is very small.  

\end{section}


\providecommand{\selectlanguage}[1]{}
\renewcommand{\selectlanguage}[1]{}

\bibliography{local_fission}